\documentclass[aps,prl,twocolumn,showpacs,groupedaddress]{revtex4-1}
\usepackage{graphicx}
\usepackage{color}
\begin{document}

\preprint{}

\title{Texture-induced modulations of friction force: the fingerprint effect}

\author{E. Wandersman}
\author{R. Candelier}
\author{G. Debr\'{e}geas}
\author{A. Prevost}
\email[]{alexis.prevost@lps.ens.fr}
\affiliation{Laboratoire Jean Perrin, Ecole Normale Sup\'erieure, UPMC Univ. Paris 6, CNRS FRE 3231, 24 rue Lhomond, 75005 Paris, France}

\date{\today}

\begin{abstract}
Dry solid friction is often accompanied by force modulations originating from stick-slip instabilities. Here a distinct, quasi-static mechanism is evidenced leading to quasi-periodic force oscillations during sliding contact between an elastomer block, whose surface is patterned with parallel grooves, and finely abraded glass slides. The dominant oscillation frequency is set by the ratio between the sliding velocity and the period of the grooves. A mechanical model is proposed that provides a quantitative prediction for the amplitude of the force modulations as a function of the normal load, the period of the grooves and the roughness characteristics of the substrate. The model's main ingredient is the non-linearity of the friction law. Since such non-linearity is ubiquitous for soft solids, this ``fingerprint effect'' should be relevant to a large class of frictional configurations and might in particular have important consequences in human (or humanoid) active digital touch.
\end{abstract}


\maketitle
Surface texture engineering by nano- or micro-patterning techniques has proven to be an efficient tool to tune frictional or adhesive properties at solid/solid~\cite{RandJAP09, BennewitzJPCM08, WuBavouzetPRE10} or solid/liquid interfaces~\cite{Bocquet}. Many of the existing designs are directly inspired by natural surfaces. Super-hydrophobic surfaces for example, reproduce the microscopic pattern observed on lotus flower leaves~\cite{Lotus}. Surfaces with a shark skin-like texture provide significant drag reduction in turbulent flows ~\cite{Sharkskin}. Geckos feet fibrillar structure provides them with remarkable adhesion capabilities, and has allowed the design of functional adhesives~\cite{Gecko}.\\
\indent In comparison, the effect of micro-patterning in solid/solid friction has so far attracted much less attention. In a recent paper however, we suggested that epidermal ridges - the regular pattern characteristics of humans and primates glabrous skin surface - strongly modify the skin internal stress elicited upon actively rubbing the fingertip over a finely abraded surface, thus enhancing tactile sensitivity~\cite{ScheibertScience09}. More precisely, such patterns were shown to produce a quasi-periodic modulation of the texture-induced subcutaneous stress field at a frequency set by the ratio between the scanning velocity and the pattern period. This finding led to the integration of fingerprint-like structure in several robotic hand designs~\cite{Robotics1, Robotics2}. Interestingly, this effect not only shows up in local stress measurements but is also observed in the global friction force signal measured while scanning a human finger on a rough surface~\cite{PrevostCIB2009}. Since the friction force signal conveys a large part of the tactile information for texture discrimination, as recently evidenced in psychophysical assays~\cite{Hayward2011}, it is crucial from both biological and robotic point of views to understand the physical origin of such texture-induced friction force modulations and to identify the parameters which control their spectral characteristics.\\
\begin{figure}
\includegraphics[width=8.6cm]{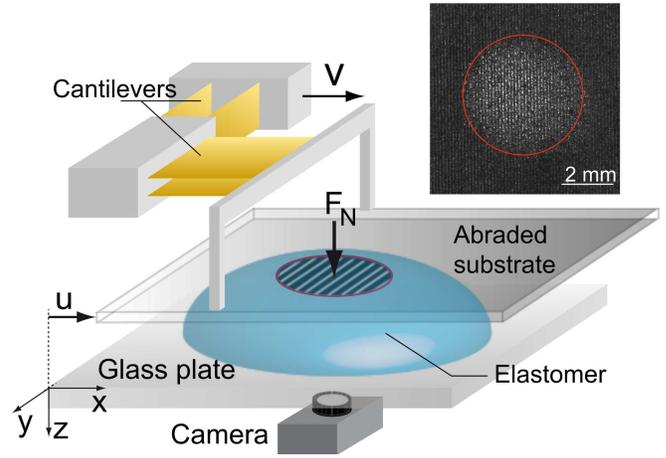}%
\caption{(color on line) Sketch of the experimental setup. Inset: image of a typical contact at $F_N=0.5$~N with $\lambda=125~\mu$m. The red circle defines the apparent contact perimeter.}
\label{Fig1}
\end{figure}
\indent To address this question, we investigate in this Letter the friction dynamics between a patterned elastomer block and a finely abraded glass surface in a sphere-on-plane geometry (Fig.~\ref{Fig1}). The elastomer blocks (thickness $\approx$18 mm) were obtained by molding liquid PolyDiMethylSiloxane-crosslinker mixture (PDMS, Sylgard 184, Dow Corning) in concave spherical lenses of radius of curvature  $R=128.8$ mm. Prior to molding, the lenses surface was patterned with a periodic square grating of depth 40 $\mu$m and spatial period $\lambda$ = 125, 218 or 760 $\mu$m (width $\lambda$/2) using soft photolithography techniques~\cite{ScheibertScience09}. The elastomer Young's modulus was measured to be $E = 2.2 \pm 0.1$~MPa. The abraded planes consisted in microscope glass slides sandblasted or abraded using a Silicon Carbide powder-water mixture. The surface topography was measured with an optical profilometer (M3D, Fogale Nanotech) from which the root mean-squared height $h_{rms}$ and surface profile autocorrelation spectrum $C(q)$ were computed~\cite{PerssonJPCM08}. Two substrates were considered, referred to as ``rough-'' ($h_{rms}\approx2.3~\mu$m) and ``rough+'' ($h_{rms}\approx5~\mu$m), respectively. For both, $C(q)$ displays a self-affine behavior as usually observed for real surfaces~\cite{PerssonJCP01} of the form $C(q)=C_0 \left(1+(q/q_c)^2\right)^{-\alpha}$ with $(C_0, q_c,\alpha)=(1.3 \, 10^{-22} ~$m$^4$, $6 \, 10^4$~m$^{-1}, 1.29)$ and $(1.3 \, 10^{-20}$~m$^4, 8.5 \, 10^3$~m$^{-1}, 1.32)$ for substrate rough- and rough+, respectively.
Substrates were rubbed against the elastomer surface along the $x$ direction at constant velocity $v$ (range 0.01~--~0.5~mm.s$^{-1}$) with a motorized translation stage (LTA-HL, Newport), under constant normal force $F_N$ (range 0.02~--~2N). $F_N$  and tangential force $F_S$ were measured with a 1 mN accuracy at 1 kHz by monitoring the deflections of two orthogonal cantilevers (Fig.~\ref{Fig1}) using capacitive position sensors (MCC-10 and MCC-20, Fogale Nanotech). The apparent contact zone, imaged in a light-transmitted geometry (Fig.~\ref{Fig1}) had a diameter ranging from 1 to 6 mm, much larger than the substrate roughness characteristic scale.

\begin{figure}
\includegraphics[width=8.6cm]{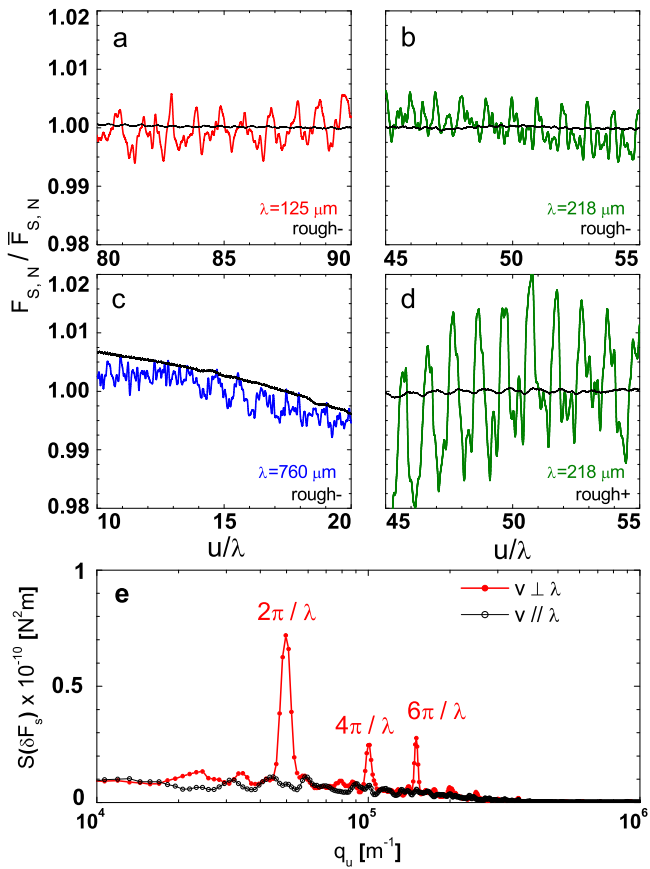}%
\caption{(color on line) (a)--(d) Blow up of $F_S/\bar{F_S}$ fluctuations (resp. $F_N/\bar{F_N}$ in thin black lines) versus $u/\lambda$ at $F_N$=0.5 N for different $\lambda$ and both substrates in steady sliding. (e) Averaged power spectra of $F_S$ fluctuations ($\lambda$=125 $\mu$m, $F_N$=0.5 N, ``rough-'' substrate), both with $v$ perpendicular ($\bullet$) and parallel ($\circ$) to the ridges.}
\label{Fig2}
\end{figure}

Typical $F_S$ signals in steady sliding are shown on Fig.~\ref{Fig2}a--d as a function of the scanned distance $u=v.t$ with $v=0.1$~mm.s$^{-1}$ perpendicular to the ridges direction and $F_N=0.5$~N. Minute but clearly measurable quasi-periodic oscillations are observed, with no equivalent in the normal force signals. Their amplitude weakly depends on $\lambda$ and increases with $F_N$. They are significantly more pronounced with the rough+ than with the rough- substrate (Fig.~\ref{Fig2}b and \ref{Fig2}d). The effect entirely vanishes when the ridges direction is aligned with the direction of motion. In the range of velocities explored, these modulations are independent of $v$. All experiments presented further are thus done at $v=0.1$~mm.s$^{-1}$ and $u$ is taken as the time-varying variable.

\begin{figure}[!h]
\includegraphics[width=8.6cm]{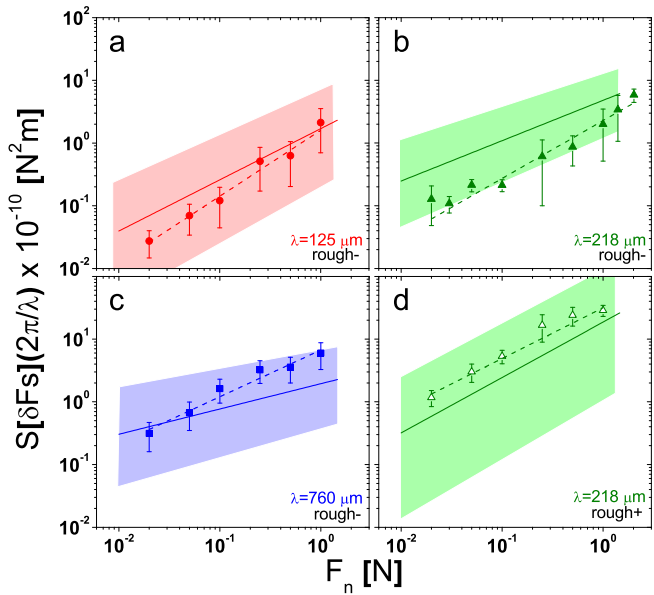}%
\caption{(color on line) (a)--(d) $S_{[\delta F_S]}(q_\lambda)$ versus $F_N$. Dashed lines are best experimental fits, solid lines are model predictions. Shaded areas bound the values of $A$ by tuning the parameters of the model.}
\label{Fig3}
\end{figure}

The power spectrum $S_{[\delta F_S]}(q)$ of $\delta F_S(u) = F_S(u)-\bar{F_S}$ (the bar stands for time averaging) was computed by averaging over a $20$~mm travel distance. All spectra exhibit well defined peaks at $q_\lambda=2\pi / \lambda$ and at corresponding harmonics (Fig.~\ref{Fig2}e). $S_{[\delta F_S]}$ at $q=q_\lambda$ increases with $F_N$ as $A F_N^{\nu}$ (Fig.~\ref{Fig3}) where values of $A$ and $\nu$ are collected in Table 1. As mentioned above, $A$ depends weakly on $\lambda$ and is significantly higher for the rough+ than for the rough- substrate. The exponent $\nu$ is close to $1$ for all experimental configurations. 

In parallel to the force measurements, rapid imaging of the contact zone was performed using a fast camera (Fastcam APX-RS, Photron, Japan) operating at 60 Hz. At all loads, contact occurs only at the ridges summits, thus yielding a large contrast between the top and bottom of the pattern and allowing for a precise tracking of the ridges edges with $\approx 10~\mu$m accuracy. No stick-slip motion of the ridges was observed. Thus, the measured force oscillations cannot be accounted for by periodic stick-slip events whose frequency is set by the period of the pattern as reported in ~\cite{BennewitzJPCM08}. This is further supported by the observation that the fluctuations are quasi-static ($v$ independent).\\

\par In the rest of this Letter, we aim at understanding the physical origin of the observed friction force modulation and the dependence of its amplitude with the normal load, pattern period and substrate roughness characteristics. We first consider a perfectly smooth elastomer surface (no patterning) rubbed against an abraded glass slide. We denote $p(x,y,u)$ and $\tau(x,y,u)$ the normal and tangential stress field at a given position $u$ of the substrate, such that $F_N(u)=\int{ p(x,y,u)\,dx \,dy}$ and $F_S(u)=\int{ \tau(x,y,u)\,dx \,dy}$. Under a local Amontons-Coulomb friction law assumption \textit{i.e.} $\tau(x,y,u)=\mu_0 p(x,y,u)$ with $\mu_0$ a uniform (material dependent) friction coefficient, the friction force reads $F_S(u)=\mu_0 F_N$ and is thus expected to be time-independent. We therefore hypothesize a weakly non-linear relationship between  $p(x,y,u)$ and $\tau(x,y,u)$ which is equivalent to postulating a pressure dependence of the local friction coefficient, \textit{i.e.} $\mu=\tau/p=\mu(p)$. The friction force now reads
\begin{equation}
\label{FsEq}
F_S(u)=\int{ \mu(p) p(x,y,u)\,dx \,dy}
\end{equation}
\noindent The spatial variations of $p(x,y,u)$ can be decomposed into a time-averaged component $\bar{p}(x,y)$ set by the macro-scale plane-on-sphere contact geometry and a time-fluctuating component $\delta p(x,y,u)$ associated with the microscopic roughness of the glass slide. For simplicity, we further assume that the elastomer is in intimate contact with the glass surface. This assumption is expected to fail at the micro-asperities scale but should be valid at intermediate scale such as the inter-ridge distance $\lambda$. In this limit, the texture-induced pressure modulations are set by the topography of the glass substrate and are independent of the local mean pressure imposed by the Hertzian geometry $\bar{p}(x,y)$. The pressure modulation field  $\delta p(x,y,u)$ is thus a sole function of the position of the substrate with respect to the contact zone $(u-x,y)$, so that the pressure field reads
\begin{equation}
p(x,y,u) = K(u) \left(\bar{p}(x,y)+\delta p(u-x,y)\right)
\end{equation}
\noindent  The normalization factor $K(u)$ ensures that $F_N$ remains constant during sliding. This condition imposes 
\begin{eqnarray}
K(u) & = & \left(1+\frac{1}{F_N}\int{ \delta p(u-x,y)\,dx \,dy}\right)^{-1} \nonumber \\
& \approx & 1-\frac{1}{F_N}\int{ \delta p(u-x,y)\,dx \,dy}
\end{eqnarray}
The latter expansion is valid when the contact diameter is much larger than the texture scale. Indeed, in this limit, the integral of the roughness-induced pressure field becomes vanishingly small with respect to the confining force $F_N$.  Rewriting Eq.~\ref{FsEq}, one obtains
\begin{eqnarray}
\label{Model}
\delta F_s(u)=F_s(u)-\bar{Fs}=\int{ (\mu(\bar{p})-\left< \mu \right>)\delta p(u-x,y)\,dx \,dy} \nonumber
\end{eqnarray}
\noindent where $\left< \mu \right>$ is defined as the ratio $\bar{F_S}/F_N$. 

The friction force fluctuations thus appear as the convolution product of a function characterizing the friction coefficient spatial heterogeneities and the texture-induced pressure modulations field. The presence of regular ridges at the surface of the elastomer imposes that the stress between ridges vanishes. This can be accounted for, in first approximation, by introducing a $\lambda$-periodic Heavyside function $H(x)=\theta(sin(2\pi x/\lambda))$ under the integral, which directly results in a spectral selection of the associated spatial mode.

\begin{figure}[htbp]
\includegraphics[width=8.6cm]{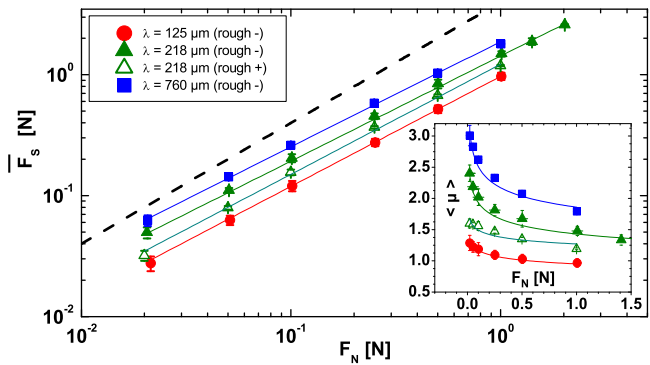}
\caption{(color on line) Time-averaged friction force $\bar{F_s}$ versus $F_N$. Different symbols correspond to different configurations. Inset: Average dynamic friction coefficient $\bar{F_S}/F_N$ versus $F_N$. On both graphs, lines are power law fits with $\gamma = 0.87 \pm 0.04$. The dashed line has a slope of 1.}
\label{Fig4}
\end{figure}

In Fourier space~\cite{PerssonJPCM08}, the fluctuating component of the friction force thus reads 
\begin{eqnarray}
\label{Modeleq}
S_{[\delta F_S]}(q_u)=  (2\pi)^4 \int{dq_v\left|\mathcal{F}\left(H(x)(\mu(x,y)-\left< \mu \right>)\right)\right|^2 S_{[\delta p]}} \nonumber
\end{eqnarray}
\noindent Wave vector $q_u$ (resp. $q_v$) refers to the parallel (resp. perpendicular) direction to the sliding direction.\\
\indent Under the assumption of intimate contact, $S_{[\delta p]}$ can be simply expressed as a function of the roughness spectrum $C(q)$ ~\cite{PerssonJPCM08} as
$S_{[\delta p]}(q)=\left(E^*/2\right)^2q^2 C(q)$, where $q$ is the norm of $(q_u,q_v)$ and $E^*$ the reduced Young's modulus. The friction coefficient spatial field, on the other hand, is estimated from global force measurements. For all experimental configurations, a power-law dependence is observed between both force components $\bar{F_S}=B.F_N^\gamma$ with $\gamma=0.87\pm0.04$ (Fig.~\ref{Fig4}).  We postulate a power-law relationship between the local shear stress and the local pressure, $\tau(x,y) = \beta \bar{p}^{m}(x,y)$ \cite{Scholz}. Assuming that $\bar{p}(x,y)$ follows a Hertz profile, the friction force $F_S$ can be derived by analytically integrating the latter equation over the contact area. This procedure yields approximated values for $\beta$ and $m$ as function of $B$ and $\gamma$. Figure \ref{Fig4} shows a comparison between the $\bar{F_S}$ versus $F_N$ obtained experimentally and through integration of $\tau(x,y) =\beta \bar{p}^{m}(x,y)$ over the contact area.

\begin{table}
\caption{\label{Table1}: Experimental and model parameters. }
\begin{ruledtabular}
\begin{tabular}{c||c|c||c|c}
Configuration & $A^{exp}$ & $\nu^{exp}$ & $A^{model}$ & $\nu^{model}$\\
\hline
$\lambda=$125 $\mu$m, rough- & 1.6 10$^{-10}$ & 1.1& 1.7. 10$^{-10}$ & 0.87\\
\hline
$\lambda=$218 $\mu$m, rough- & 2.3 10$^{-10}$ & 0.92& 4.8 10$^{-10}$ & 0.64\\
\hline
$\lambda=$218 $\mu$m, rough+ & 3.2 10$^{-9}$ & 0.81& 1.9 10$^{-9}$ & 0.88\\
\hline
$\lambda=$760 $\mu$m, rough- & 6.7 10$^{-10}$ & 0.75& 2.0 10$^{-10}$ & 0.4\\

\end{tabular}
\end{ruledtabular}
\end{table}

Both predicted and measured amplitudes of $S_{[\delta F_S]}(q_\lambda)$ with $F_N$ are presented on Fig.~\ref{Fig3}, for all experimental configurations, using the set of parameters ($\beta$, $m$, $C_0$, $q_c$, $\alpha$) obtained as described earlier. The predicted amplitude of $F_S$ modulations is in reasonable agreement with the data, without any adjustable parameter. The model captures in particular the analytic increase of $S_{[\delta F_s]}(q_\lambda)$ with $F_N$, with an exponent $\nu$ comparable to its experimental counterpart (see Table 1 and Fig.~\ref{Fig3}).  By tuning systematically the input of the model ($m$, $\lambda$), $\nu$ is found to increase linearly with $m$ and to decrease linearly with $\lambda$ in the investigated range (not shown).  The large uncertainty of $\nu$ ($\sim$ 0.1)  in the model is thus mostly due to the experimental inaccuracy in evaluating the parameter $m$. The predicted amplitude $A$ is trivially proportional to $C_0$, slightly decreases with $\lambda$ and falls rapidly to zero when $m$ approaches unity, \textit{i.e.} when Amontons-Coulomb friction's law is verified locally.

This study shows that any non-linearity in the friction law leads to the development of texture-induced friction force fluctuations. Since purely linear friction laws are scarcely observed, the present mechanism should be relevant to most practical situations. However, such fluctuations are expected to be in practice hardly detectable when the fixed substrate (here the PDMS block) is smooth and the contact zone diameter is much larger than the typical roughness scale. In contrast, the presence of a regular pattern at the surface of the block operates a spectral filtering of the texture induced pressure modulation which induces a mode selection of the force fluctuations at a particular frequency. The amplitude of these force fluctuations can then become an order of magnitude larger than in the smooth case as shown by the present study. This mechanism was here examined for the simplest possible pattern, \textit{i.e.} regular parallel stripes, but richer force signal spectra should be similarly obtained with more complex patterns.  

In essence, the macroscopic friction force signal carries information about the spectral content of the microscopic substrate topography at frequencies defined by the elastomer micro-pattern. A practical consequence is that $F_S$ fluctuations amplitudes could be used to discriminate surfaces having small differences in their roughness at scales much smaller than the contact extension. Hence, under the same conditions, force fluctuations are roughly 10 times larger for the substrate having a $h_{rms}\approx 5 \mu$m than for the one with $h_{rms}\approx 2.3 \mu$m. This mechanism could also facilitate the detection of motion. The transition between static and sliding contact should manifest as the sudden appearance of large single mode oscillations of the friction force. One may finally foresee the possibility to evaluate the scanning velocity, based on the sole friction force signal, by continuously extracting the peak frequency of these oscillations.  We believe that this ``spectrotribometry'' approach may be relevant to human digital touch and prehension, owing to the presence of fingerprints, and that it could be easily implemented in tactile robotic sensing devices. 

\begin{acknowledgments}
The authors deeply thank F. Petrellis for illuminating mathematical surgery, F. Zalamea for his valuable help, A. Chateauminois and C. Fr\'etigny for fruitful discussions, and acknowledge financial support from ANR-DYNALO NT09-499845.
\end{acknowledgments}

\end{document}